# ANNUAL REVIEWS

*Annual Review of Nuclear and Particle Science*

# Detection and Calibration of Low-Energy Nuclear Recoils for Dark Matter and Neutrino Scattering Experiments

Jingke Xu,[1] P.S. Barbeau,[2] and Ziqing Hong[3]

[1]Lawrence Livermore National Laboratory, Livermore, California, USA; email: xu12@llnl.gov

[2]Department of Physics, Duke University, Durham, North Carolina, USA

[3]Department of Physics, University of Toronto, Toronto, Ontario, Canada

## Abstract

Detection of low-energy nuclear recoil events plays a central role in searches for particle dark matter interactions with atomic matter and studies of coherent neutrino scatters. Precise nuclear recoil calibration data allow the responses of these dark matter and neutrino detectors to be characterized and enable in situ evaluation of an experiment's sensitivity to anticipated signals. This article reviews the common methods for detection of nuclear recoil events and the wide variety of techniques that have been developed to calibrate detector response to nuclear recoils. We summarize the main experimental factors that are critical for accurate nuclear recoil calibrations, investigate mitigation strategies for different backgrounds and biases, and discuss how the presentation of calibration results can facilitate comparison between experiments. Lastly, we discuss the challenges for future nuclear recoil calibration efforts and the physics opportunities they may enable.

## Contents

# 1. INTRODUCTION

Searches for dark matter and neutrino interactions are at the forefront of particle physics experimentation. These efforts aim to address several profound questions facing physicists today, including the composition of the universe and its evolution (1), the origin of the matter–antimatter asymmetry that led to our very existence (2), and whether the Standard Model (SM) of particle physics, which explains nearly everything about known particles, can continue guiding us into uncharted territories of experimental particle physics (3).

Of all known particles, neutrinos have one of the weakest interaction strengths, with a cross section of $10^{-40}$ cm² or lower for their interactions with electrons and protons. Interactions of dark matter particles with matter have not been definitively observed to date, and the interaction strength is expected to be similar to, or weaker than, that of neutrinos. These extraordinarily low-cross-section interactions require detection experiments to have extremely low background rates, large detector sizes, and strong signal–background discrimination capabilities.

In direct detection dark matter experiments, the main signal channel being pursued is elastic dark matter scattering off nuclei. In the simplest case, a dark matter particle such as a weakly interacting massive particle (WIMP) with galactic speed ($\sim10^{-3}c$) interacts coherently with all nucleons inside a nucleus and transfers a fraction of its kinetic energy to the recoiling nucleus. This interaction is analogous to the recently observed coherent elastic ν-nucleus scatter (CEνNS) process (4, 5). In addition to the coherent boost of the interaction rate, the nuclear recoil nature of the expected dark matter and neutrino CEνNS signals also allows for rejection of ambient radiation backgrounds. The dominant background in a typical radiation detector is from cosmic rays and gamma rays, which primarily produce recoiling electrons, so detector technologies that can distinguish nuclear recoils from electron recoils have significant advantages in the experimental pursuits of these low-cross-section physics.

Searches for dark matter– and neutrino-induced nuclear recoil signals have proven to be a significant challenge. Many dark matter models predict recoil signals at the keV level for commonly used targets including argon, silicon, germanium, and xenon (6). The CE$\nu$NS process for common neutrino sources produces signals in a similar energy region, which becomes even lower for low-energy neutrinos or heavy targets. These nuclear recoils are inefficient at producing detectable signals, such as scintillation photons or ionization electrons, compared with electron recoils (7), making them even more difficult to observe. As a result, the CE$\nu$NS process was only first observed experimentally in 2017 (5), more than 40 years after it was proposed (4). WIMP dark matter interactions remain to be observed.

As nuclear recoil–based dark matter and neutrino experiments continue to improve their sensitivities, the need to accurately determine their signal responses has grown. This is especially true for low-mass dark matter searches (8) and reactor CE$\nu$NS detection (9, 10), both of which require detection of sub-keV nuclear recoils. The production and detection of such low-energy recoils pose a series of new challenges and have begun to be explored only in the last few years in silicon (11, 12), germanium (13–16), and xenon (17–19). At higher energies, although calibration data are more abundant, different experiments often report conflicting results, which remain the main sources of uncertainty for current CE$\nu$NS studies. These challenges call for a systematic investigation of the commonly practiced calibration and analysis techniques.

This article reviews the widely used nuclear recoil detection and calibration techniques. We examine the critical elements for a successful calibration experiment, discuss the general principles to be followed in the analysis of calibration data, and make recommendations on how experimental results can be reported to facilitate comparison between efforts. We aim to provide useful guidance for future calibration endeavors to meet the increasingly demanding calibration needs of ongoing dark matter and neutrino experiments.

## 2. NUCLEAR RECOIL SIGNALS IN PARTICLE EXPERIMENTS

### 2.1. WIMP Dark Matter Search

The existence of dark matter is strongly supported by a range of cosmological observations (1), from rotation curves of galaxies to the gravitational lensing and X-ray imaging of colliding galaxy clusters and from the cosmic microwave background (CMB) radiation to the big bang nucleosynthesis. Approximately five times more nonluminous (or dark) matter than baryonic matter is required to reconcile the Lambda cold dark matter ($\Lambda$CDM) model with the observed CMB power spectrum (20). Dark matter is widely believed to consist of exotic particles beyond the SM, and if they interact with SM particles, the interaction strength would be at the weak scale or even lower.

WIMPs are leading dark matter candidates (21). For WIMPs in the mass range of GeV/$c^2$ to TeV/$c^2$, thermal relics from the big bang can naturally explain the required dark matter density today if they interact with SM particles at the weak scale, which has been commonly referred to as the WIMP miracle (22, 23). Various extensions to this simple WIMP dark matter model have also been proposed (24, 25).

For a WIMP mass of $M_\chi$ and target mass of $M_T$, the maximum dark matter energy transfer from the WIMP to an at-rest target in an elastic scatter is

$$E_R \leq E_\chi \frac{4M_T M_\chi}{(M_T + M_\chi)^2} \qquad 1.$$

and the maximal energy transfer occurs when the target mass and the WIMP mass are comparable (6). With a 100 GeV/$c^2$ WIMP and a 100 GeV/$c^2$ target, the maximum nuclear recoil energy is around 50 keV.

WIMPs may interact with both electrons and nuclei. When a slow WIMP scatters off shell electrons of an atom, the momentum transfer is dominated by the electron binding energy, and the typical energy transfer is of order 1 eV (26). This is below the thresholds of most particle detectors today, and the expected interaction rate is also low without a coherent enhancement. These factors have led to the dominance of nuclear recoil–based WIMP search efforts today. One may construct couplings that eliminate dark matter–nucleon interactions at the tree level, thus making dark matter particles electrophilic. However, effective dark matter–nucleon interactions can reemerge at the one-loop level with appreciable cross-section values (27). Therefore, detection of low-energy nuclear recoil signals remains important even for some exotic dark matter models.

Direct detection WIMP search efforts made substantial progress in the past few decades and have ruled out large parameter spaces for the benchmark spin-independent and spin-dependent WIMP–nucleon interactions with a standard halo assumption (28). A generalized treatment of WIMP–nucleon interactions is the effective field theory (EFT) approach (24). The EFT framework considers a broader range of interaction modes that preserve gauge invariance and introduces over a dozen new effective operators. In these new modes, WIMP-induced nuclear recoil spectra often deviate from those predicted by the simple benchmark models, and some interactions can lead to relatively suppressed event rates at low energies while the higher-energy interactions are largely preserved. Furthermore, dark sector dark matter models can also predict dark matter interactions at energies below the currently pursued keV level (25). These developments have led direct detection experiments to expand their dark matter search energy regions of interest down to sub-keV levels and up to hundreds of keVs (29–35). As more dark matter interaction parameter space is being explored, precise signal calibrations in these new energy regions are required.

## 2.2. CEνNS Detection

Neutrinos are SM particles, but their properties cannot be fully described by the SM (36). Most notably, neutrinos of all flavors are predicted to have zero mass, but the observation of neutrino oscillations between flavors requires their masses to take on finite values (37–39). In addition, the current upper limits on neutrino mass (40) are orders of magnitude smaller than the masses of other SM particles. This anomaly may be explained by the seesaw mechanism with an additional Majorana mass term, which requires a new heavy right-handed neutrino with no weak force interaction (sterile neutrino) to accompany the light left-handed neutrinos. Experiments studying neutrino properties can thus provide a pathway to search for new physics beyond the SM.

Neutrinos are usually detected through their elastic scatters with electrons and inelastic interactions with nucleons. They typically have a cross-section value of $10^{-40}$ cm$^2$ or lower. These channels lose sensitivities to neutrinos below a few MeV of energy due to either a diminishing cross section or irreducible backgrounds. On the other hand, the flavor-blind CEνNS process has a relatively high interaction cross section of $\sim 10^{-39}$–$10^{-38}$ cm$^2$ (4), and like neutrino–electron scatters, CEνNS is a pure kinematic interaction and does not have an intrinsic energy threshold. Therefore, detectors sensitive to low-energy nuclear recoils induced by CEνNS can provide information about neutrinos complementary to that from charged-current detection techniques.

The interaction cross section of the CEνNS process is approximately formulated as

$$\frac{d\sigma}{dT} = \frac{G_F^2 M}{4\pi}\left(1 - \frac{MT}{2E_\nu^2}\right)Q_w^2\left[F_w(q^2)\right]^2 \qquad 2.$$

where $G_F$ is the Fermi constant, $M$ is the target nucleus mass, $E_\nu$ is the neutrino energy, $T$ is the nuclear recoil energy, $Q_w$ is the weak charge of $Z(1 - 4\sin^2\theta_W) - N$, and $F_w(q^2)$ is the weak form factor that approaches 1 at low momentum transfers. Due to the value of the weak mixing angle $\sin^2\theta_W \sim 0.23$, the CE$\nu$NS cross section is enhanced by approximately the square of the neutron number of the nucleus, explaining its relatively high value.

Despite its high cross section, the CE$\nu$NS process is extremely difficult to detect. Because of its elastic scatter nature, the final state of the nucleus is the same as the initial state, except that a minuscule kinetic energy of $T < 2E_\nu^2/(M + 2E_\nu)$ is transferred from the neutrino to the nucleus. High-energy neutrinos, such as those produced at the Spallation Neutron Source (SNS), can cause nuclear recoils up to 100 keV in argon and 30 keV in CsI (5, 41), but for reactor neutrinos with few-MeV energies, the maximum recoil energy decreases to around 1 keV or lower for the same targets (9). Moreover, the majority of this kinetic energy is dissipated as heat such that the signal observed by a scintillation or ionization detector is further quenched (7). Up to today, the CE$\nu$NS process has been measured only with high-energy neutrinos ($\sim$30 MeV) at the SNS (5, 42).

The CE$\nu$NS process is blind to neutrino flavors, and thus the measured rate includes contributions from all three active neutrino species. With sufficient precision, a CE$\nu$NS experiment, being insensitive to neutrino oscillation between the active flavors, can provide a more definitive confirmation of neutrino oscillation into a sterile state, complementing the charged-current short-baseline oscillation experiments (43). Its high cross section and flavor blindness also enhance its sensitivity to core-collapse supernovae explosions (44), and it can play a significant role in multimessenger studies of supernovae in addition to gravitational wave and electromagnetic observations (45). Practical applications of CE$\nu$NS may also include the nonintrusive monitoring of nuclear reactor and spent fuel storage sites that are important for nonproliferation projects (9, 46). Lastly, as recently demonstrated, a beam-CE$\nu$NS experiment can also be sensitive to light dark matter that may be produced by high-energy particle collisions and their subsequent decays (47, 48).

## 2.3. Neutron-Induced Nuclear Recoils

Neutrons can produce nuclear recoils through elastic and inelastic scatters with nuclei that mimic signals in WIMP dark matter or neutrino scatter experiments; as a result, neutron-induced recoils are seen as one of the most dangerous backgrounds in such rare event searches. However, neutrons also play an irreplaceable role in providing signal calibration for these nuclear recoil search experiments. As elaborated in Section 4, nuclear recoils of a desired energy and direction can be produced through interactions of neutrons with the appropriate energy and timing structure using coincidence tagging methods. Not only are these neutron-induced nuclear recoils critical for the detectors' response to be understood, they also provide a means to evaluate the signal acceptance to mitigate potential biases in rare event searches.

## 3. DETECTION METHODS FOR NUCLEAR RECOILS

Although neutrinos and dark matter particles are considered invisible, recoiling nuclei produced by their interactions inside a medium can lead to observable signals. Measurement of ionization signals and/or detection of photon emission following deexcitation and recombination of electron–ion pairs can enable the primary interactions to be reconstructed and the properties of the incoming particles to be deciphered. In addition to atomic interactions, slow nuclear recoils dissipate a large fraction of their kinetic energy through vibration of atoms. Therefore, a heat-sensitive detector such as a low-noise bolometer has the potential to probe into the very-low-energy region of these rare interactions.

This section reviews the energy dissipation mechanisms of nuclear recoils in a medium and how the resulting signals can be collected. Such detector characteristics also govern a detector's calibration requirements and its achievable experimental sensitivities.

## 3.1. Energy Loss of Recoiling Nuclei in a Medium

A recoiling nucleus loses its kinetic energy through a cascade of collisions with nearby atoms in the medium. Due to the large mass disparity between a nucleus and an electron, the energy transfer from a slow-moving nucleus to shell electrons is inefficient, and thus only a small fraction of the energy is channeled into electronic states such as atomic ionization and excitation. The majority of the energy is transferred to atomic motion, which increases the temperature of the medium. The fraction of the initial nuclear recoil energy transferred to electronic states is important for detector technologies that sense only ionization and scintillation signals. In addition, this value is usually different for nuclear recoils, electron recoils, and other backgrounds and thus serves as a discriminator between signals and backgrounds in an experiment that detects both heat and electronic signals.

Different approaches have been studied to predict the electronic yield of nuclear recoils in matter. The popular Stopping and Range of Ions in Matter (SRIM) software package is used to numerically calculate the transport of ions in matter (49). Analytically, the most widely used model is the one developed by Lindhard et al. (50–52). Using simplified cross sections for the interactions of electrons and nuclei, this model calculates the cascade energy losses to different dissipation channels. It relies on a list of assumptions, including that

- electronic and nuclear interactions can be treated separately,
- the kinetic energy is mostly transferred to atomic motion during the cascade,
- the energy gained by a nucleus through collisions is small compared with the energy of the projectile nucleus,
- the atomic binding energies of electrons are negligible, and
- electrons do not cause appreciable nucleus motion.

With these assumptions, Lindhard et al. obtained the fraction of nuclear recoil energy transferred to electronic states when the projectile nucleus is of the same kind as the target nuclei,

$$f = \frac{kg(\epsilon)}{1 + kg(\epsilon)}, \qquad\qquad 3.$$

where $k$ is the electronic stopping power coefficient and $g$ is a parameterized function of the reduced dimensionless energy $\epsilon$.

The Lindhard theory successfully describes the measured ionization yields in different detector media across a large energy range and has also been tested for the total electronic yield in detectors that measure both scintillation and ionization signals (53–55). In addition to predicting the average yield, this theory also calculates the intrinsic variance of the $f$ factor (50), which is analogous to a Fano factor for electron recoils and may be evaluated experimentally (56). This additional spread may cause observed signals to deviate from a simple Gaussian or Poisson distribution and poses a challenge to the accurate modeling of low-energy nuclear recoil signals in experimental efforts that are sensitive to this effect.

As detector technologies improve, experiments have started to explore energy regions where some of the Lindhard assumptions are no longer valid, including but not limited to the neglect of electron-binding energies. Recent works by Sorensen (57) and Sarkis et al. (58) relax this assumption and predict a roll-off of scintillation and ionization yields below certain threshold energies. These new theoretical developments have begun to be tested experimentally.

In detectors that collect only scintillation photons, the scintillation yield per path length can be empirically described with Birks' law,

$$\frac{\mathrm{d}L}{\mathrm{d}x} = \frac{S}{1 + kB\frac{\mathrm{d}E}{\mathrm{d}x}}\frac{\mathrm{d}E}{\mathrm{d}x}, \qquad\qquad 4.$$

where $S$ is the scintillation efficiency, $dE/dx$ is the linear energy transfer by the ionizing particle to the medium, and $kB$ is the Birks quenching coefficient that reduces the observed scintillation yield for high ionization densities (59). This approximation was originally developed for organic scintillators but was later found to be applicable to some inorganic scintillators as well. Mei et al. (55) combined the Lindhard theory and Birks' law and demonstrated some success in describing the observed scintillation yields of noble liquids.

## 3.2. Subdominant Energy Loss Channels

The energy loss treatment described in Section 3.1 assumes that a recoiling nucleus comoves with the atomic shell electrons immediately after the initial impact. But as pointed out by Migdal (60) in 1941, the electron cloud may be displaced relative to the nucleus by the sudden momentum transfer, and as a result, excitation and ionization can occur following the production of a nuclear recoil. These atomic excitation and ionization losses are a result of energy transfer from the incoming particle to the atom directly impacted and thus are distinct from those produced by the recoiling nucleus moving through the medium. Ibe et al. (61) reformulated this phenomenon in 2017 by coherently treating the energy distribution between atomic electrons and the nucleus during the atomic recoil process, which ensures the conservation of energy, momentum, and interaction probability.

When this effect is considered, a WIMP dark matter particle or neutrino scatters coherently with an atom, and the whole center-of-momentum energy is available for producing atomic excitation and ionization. Therefore, interactions, which would otherwise produce only low-energy nuclear recoils, have a finite probability to produce significant electronic energy depositions. Despite the low probability, this higher electronic energy, along with its ability to be detected more efficiently than nuclear recoils, has led to the promise of new low-mass dark matter sensitivities even in relatively high-threshold detectors. It was further demonstrated that for dark matter masses of $\mathcal{O}(100)$ MeV and a heavy mediator that couples comparably with nucleons and electrons, the Migdal-induced electronic energy deposition can dominate over direct dark matter–electron scatters (26, 62). Several experiments have used this effect to claim improved sensitivities to dark matter masses below the GeV level (29, 63–65), pending experimental verification of the process.

When a nucleus moves inside a crystallographic structure, its energy loss to atomic collisions can be substantially suppressed if its direction is aligned to the crystal axes or planes (66). As a result, its range can be extended and a bigger fraction of its energy may be transferred to atomic excitation and ionization when compared with that in amorphous materials. This effect was proposed to reconcile conflicting observations in dark matter experiments using crystalline and amorphous materials (67) but has yet to be confirmed in a calibration experiment. If observed, this directional effect may be leveraged to enhance sensitivities of dark matter and CE$\nu$NS experiments by boosting the signal strength and suppressing backgrounds using directional information.

## 3.3. Nuclear Recoil Detection Technologies

Scintillation and ionization detection techniques have been under continuous development for over a century, leading to their relative maturity today. As a result, although only a small fraction of the nuclear recoil energy from dark matter and neutrino scatters is dissipated in the electronic channels, experiments focusing on scintillation and ionization detection have led these rare event searches. Inorganic scintillators that can be purified aggressively and achieve low background rates, such as NaI(Tl) and CsI(Na) crystals (48, 68), and noble element detectors, including liquid argon and xenon (69, 70), are widely used. Some detector media can produce different scintillation pulse shapes for nuclear recoils and electron recoils, providing background rejection capabilities (69, 71).

However, given a typical scintillation $W$-value (the energy required to produce a single quantum) of 100–200 eV for nuclear recoils and a light collection efficiency of $\mathcal{O}(10\%)$ in most detectors, scintillation-based experiments often report energy thresholds of several keV or higher for nuclear recoils.

Ionization detectors using germanium (31, 72, 73), silicon (74), and noble elements (71, 75–77), on the other hand, can achieve lower energy thresholds thanks to their $\mathcal{O}(100\%)$ signal collection efficiencies and, in the case of semiconductors, lower ionization $W$-values. A detector may directly read out the charge signals through high sensitivity charge amplifiers or amplify the ionization signals through electroluminescence (75, 76), field multiplication (78), or Neganov–Luke phonon generation (31). These technologies are highly complementary to each other due to their different ionization $W$-values, different achievable active masses, different demonstrated background levels, and background rejection capabilities.

Since the majority of nuclear recoil energy in rare event detectors goes into the heat channel, a low-background bolometer can have the sensitivity to recoil signals of very low energies. Detectors using germanium, silicon, and $CaWO_4$ crystals have demonstrated sensitive bolometric readout with cryogenic sensors such as transition edge sensor and neutron transmutation doped thermistors (72, 74, 79). These detectors need to operate at mK-level temperatures to reduce thermal phonon backgrounds and to improve signal collection performance. A bolometer can read out either athermal phonons or thermal phonons. Athermal phonon detectors are sensitive to only part of the phonon population but benefit from a faster detector response, usually with a timing capability of microseconds. Thermal phonon detectors, on the other hand, can sense most of the recoil energy but have slower responses (milliseconds). In both designs, resolutions of a few eV have been demonstrated with gram-level detectors (80, 81). New sensor technologies are also being developed, including metallic magnetic calorimeters (82) and microwave kinetic inductance detectors (MKIDs) (83). The MKIDs benefit from their intrinsic capability of multiplexing via radio-frequency superconducting quantum interference device readouts. Bolometric sensors themselves may also function as detector targets because they are sensitive to energy depositions well below the eV level. For example, dark matter limits were obtained using merely 4.3 ng of superconducting nanowires (84). The sensitivity reach of this approach is limited by the small mass of sensing devices, which, if increased, often results in reduced energy resolutions and increased noises (85).

A special type of heat-sensitive nuclear recoil detection uses phase changes. In a bubble chamber, the liquid medium is overheated to above the boiling temperature but is kept from boiling due to the lack of nucleation centers; when interactions with a high-energy deposition density, such as nuclear recoils, occur, they can create a critically sized bubble and seed the formation of observable bubbles that enable the nuclear recoil events to be recorded (86). By tuning the thermal conditions of the detector, a bubble chamber can be insensitive to electron recoils and record only nuclear recoils, giving them significant advantages in nuclear recoil–based rare event searches. Although bubble formation is typically a slow process, high-speed cameras and acoustic sensors can enable the interaction time to be recorded. It was also recently demonstrated that neutron-induced nuclear recoils can trigger nucleation and ice forming in supercooled water (87).

A detector that records multiple signal channels for an event can have enhanced sensitivities to rare interactions. Noble liquid time projection chambers (TPCs) can collect both scintillation and ionization signals, which, when combined, enable them to have improved energy resolution, three-dimensional position reconstruction, and nuclear recoil–electron recoil discrimination down to the keV level. Some cryogenic bolometers are capable of collecting both ionization signals and heat and suppress both electron recoil backgrounds and heat-only noises. Bubble chambers with scintillation sensitivities are also being developed (88).

# 4. NUCLEAR RECOIL CALIBRATION TECHNIQUES

Any physical process that produces recoiling nuclei may be used for the calibration of nuclear recoil detectors. The most common technique involves the elastic scatter of neutrons off a target, but alternative processes such as inelastic neutron interactions, gamma-induced nuclear recoils, and even charged particle scatters can be used. These techniques are discussed in this section.

## 4.1. Neutron Elastic Scatter Calibration

A single elastic scatter between a neutron and an at-rest target nucleus transfers a kinetic energy of up to $4E_n m_n m_T / (m_n + m_T)^2$, where $E_n$ and $m_n$ are the energy and mass of the incoming neutron and $m_T$ is the mass of the target. If the scattered neutron direction is further measured, the nuclear recoil energy becomes kinematically constrained. For a laboratory neutron scatter angle of $\theta$ the kinetic energy transfer $E_T$ in this process is

$$E_T = \frac{2E_n m_n^2}{(m_T + m_n)^2} \left( \frac{m_T}{m_n} + \sin^2\theta - \cos\theta \sqrt{\left(\frac{m_T}{m_n}\right)^2 - \sin^2\theta} \right). \qquad 5.$$

This process is capable of producing nearly monoenergetic nuclear recoils and is the main method discussed in this section. A typical experimental setup is illustrated in **Figure 1**. Implementation of this method in an experiment depends on the signal energy of interest, the mass and excited energy levels of the target, the timing and other characteristics of the target detector, the properties of the available neutron source, and the expected background rates relative to that of signals.

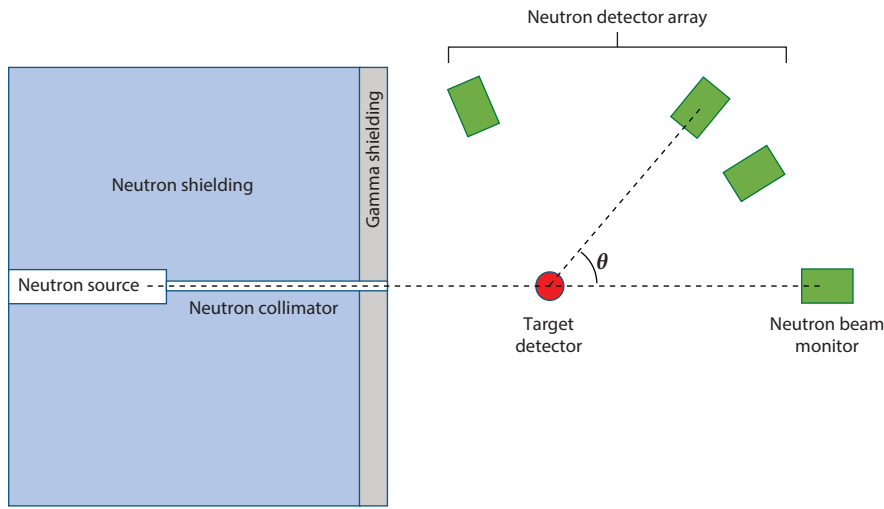

**Figure 1**

Illustration of a typical experimental setup to measure nearly monoenergetic nuclear recoils using elastic neutron scatters. Neutrons of known energy produce nuclear recoils with a well-defined energy distribution when scattering off a target at a specific angle. The neutron scatter angle is usually informed by placing detectors with gamma-neutron discrimination capabilities at fixed locations, where direct neutrons from the source should be suppressed with shielding. The neutron passage inside the shielding (collimator) may be tapered to improve beam purity, and a neutron detector may be placed right behind the target detector as a beam monitor.

To take advantage of elastic neutron scatter kinematics, a nuclear recoil calibration experiment needs the following basic elements:

1. neutrons of known energy selected to produce the desired range of recoil energies,
2. a compact target detector with minimal inactive materials to reduce neutron multiscatter probability or a detector with sufficient position sensitivity to differentiate multiscatters if a large target volume is used, and
3. the detection of outgoing neutrons.

Direct knowledge of neutron energy is the most critical element for elastic scatter neutron calibrations. Neutrons produced by some accelerator-based reactions, such as proton-lithium (p-Li) interactions or deuterium-deuterium/tritium (D-D/D-T) fusions, or by photoneutron emission sources can have well-defined energies and have been widely used in nuclear recoil calibrations. Sources that emit neutrons of continuous energy may also be used in such measurements if the energy of each incident neutron can be measured through the neutron time-of-flight (nTOF) technique. This may be achieved by measuring both the neutron production time (such as with neutrons produced by pulsed accelerators or when gammas accompany the neutron emission) and the neutron impact time in the detector (if the detector response is fast) or by allowing the neutrons to scatter once in an instrumented scatter volume before impacting the target detector (89). Combining the neutron path length and travel time can allow the neutron velocity and energy to be calculated. The accuracy of the evaluated neutron energy varies for different sources and measurement techniques, as is further discussed in Section 4.3.

Neutron scatter angles are usually obtained by tagging the scattered neutrons with dedicated neutron detectors, which have locations relative to the beam and the target detector. These detectors should be sufficiently shielded from the source neutrons such that they are sensitive only to scattered neutrons. To ensure the tagged neutron interactions are dominated by single scatters, target detectors are usually designed to be compact, which also reduces the spread of scatter angles in tagged events. If the target or tagging detector can be position sensitivity, the interaction positions can be further used to reject multiple scatter backgrounds and to improve the neutron scatter angle calculation. Inactive materials around the target and tagging detectors should be minimized to avoid additional neutron interactions. Similarly, when multiple tagging detectors are used to increase the overall coincidence tagging efficiency, the placement of the tagging detectors should be planned to mitigate neutron multiple scatters among these detectors.

The desired neutron energy and scatter angle are the first parameters to be decided while planning for a nuclear recoil calibration. The neutron energy should be chosen such that the desired recoil energy is produced by scatters at intermediate angles with an adequate rate, and the elastic scatter cross section at the chosen energy and angle should dominate that of competing processes such as inelastic interactions. Tagging shallow neutron scatters (near 0°) or extreme backscatters (near 180°) can lead to high rates of accidental coincidences between interactions inside the target detector and the detection of unrelated beam neutrons due to the close proximity of the neutron detectors to the beam. An experiment collecting >90° neutron scatter events may observe low signal rates from the reduced elastic scattering cross-section values at large angles and high background rates due to the vicinity of tagging detectors to the neutron source and its shielding.

To mitigate false coincident backgrounds, the nTOF and particle identification (PID) techniques should be used when possible. Fast neutrons with MeV-scale energy travel at a speed of ~1 cm/ns, which is substantially lower than that of gammas (~30 cm/ns). Therefore, with sufficient timing resolution, the time differences between a neutron leaving the source, impacting the target detector, and hitting the tagging detector can be measured and compared to the expected flight times to reject gamma coincidence, neutron multiple scatter background, and random co-incidence backgrounds. For target detectors with slow signals, as is the case for many bolometers and some ionization detectors, the neutron scatter time cannot be accurately measured, but timing from the neutron source, when available, can enable the full-path nTOF between the source and the tagging detector to be used. Moreover, if the tagging detector and/or the target detector can differentiate neutron and gamma interactions, false coincidence backgrounds can be further suppressed. Pulse shape discrimination is the PID technique used most often and has been extensively studied for common neutron detectors using organic and inorganic scintillators. For experiments using low-energy neutrons (<300 keV), tagging detectors may have a compromised efficiency for detecting the prompt neutron interactions or differentiating neutron events from gammas. In these cases, capture-based neutron detectors such as boron- or lithium-loaded scintillators may be used instead. In particular, Li has a relatively high fast neutron capture cross section that permits recovery of some nTOF capability. The continuous nature of the nTOF and PID parameters may also enable residual backgrounds inside the nTOF and PID regions of interest to be estimated on the basis of extrapolation from events outside these regions, so these backgrounds can be statistically subtracted.

In some experiments involving large target detectors, it may not be practical to tag the outgoing neutrons after single scatters. In these setups, nuclear recoil calibrations can be simply carried out by recording all neutron-induced interactions; the energy calibration factor may then be inferred by comparing the observed energy spectrum with that simulated using parameterized detector response models. When only the recoil response at the maximum single scatter energy is extracted, such an experiment is referred to as an end-point measurement. This calibration technique may suffer higher systematic uncertainties than the tagged method due to the additional model dependence, but it can provide valuable data for in situ signal efficiency evaluations. In large target detectors with sufficient position and timing resolutions, if the scatter vertices and the order of interactions can be determined, multiple neutron scatter interactions inside the target can also be used for neutron energy transfer calculations even without dedicated neutron tagging detectors. Preliminary success has been demonstrated in this approach using the Large Underground Xenon (LUX) detector (90).

## 4.2. Alternative Calibration Methods

Inelastic neutron scatters on nuclei have been used for nuclear recoil calibrations (91, 92). Such an interaction can produce a fixed-energy deexcitation gamma (or a conversion electron) and a low-energy nuclear recoil simultaneously. When the incoming neutron energy is selected to be right above that required to excite the nucleus, the produced nuclear recoils are nearly monoenergetic with minimal dependence on the neutron scatter angle. The nuclear recoil response in the target can then be evaluated by studying the shift or broadening of the deexcitation gamma energy peak. Using this method, Jones & Kraner (92) measured $^{73}$Ge recoils in a germanium crystal down to around 1 keV of energy in 1971.

Low-energy nuclear recoils are also produced in processes not involving fast neutrons. For example, when a thermal neutron is captured by a nucleus, one or more high-energy gammas may be emitted, which can cause the newly formed nucleus to recoil. The nuclear recoil energy is conveniently constrained by the gamma energy because the initial momentum carried by the incoming thermal neutron is negligible. When a single gamma ray of $E_g$ is emitted, the nuclear recoil energy will be $E_g^2/2M_N$, where $M_N$ is the mass of the produced nucleus. For a typical $E_g = 8$ MeV and $M_N = 100$ GeV/$c^2$, the nuclear recoil energy is around 300 eV. In reality, several gamma rays are usually emitted in such a process and the nuclear recoils will have a finite energy spread. In these experiments, some of the emitted gammas can be used to tag the gamma-induced nuclear recoil.

Jones & Kraner (93) used this method to successfully measure 0.25 keV $^{73}$Ge recoils in a germanium detector in 1974. Similar measurements have also been proposed for argon and xenon. Neutron captures on $^{40}$Ar can produce two gamma rays at 5.6 MeV and 0.5 MeV, along with a recoiling $^{41}$Ar of about 0.4 keV energy (46). In xenon, neutron captures on $^{129}$Xe and $^{131}$Xe produce nuclear recoils up to 0.3 keV (94) and can enable calibrations at lower energies than what has been demonstrated using elastic neutron scatters. A similar approach is being investigated for solid state detectors including CaWO$_4$ and silicon (95, 96).

Joshi (97) studied the feasibility of using the resonant photonuclear scattering process to further explore gamma-induced nuclear recoil calibrations. It was suggested that nuclear recoils produced by nuclear resonance fluorescence can preserve the simplicity of elastic photonuclear scattering and have enhanced interaction cross sections. More importantly the resonant photonuclear scattering process can access a wider range of nuclear excitation levels than the neutron capture method and also mitigate broadening of the nuclear recoil energy from additional gamma emissions. This method has not been experimentally tested to our knowledge.

More exotic calibration techniques may involve charged particles. The COUPP collaboration used negative pions of 12 GeV/$c$ to calibrate the nucleation efficiency of iodine recoils in a CF$_3$I bubble chamber (98). This approach was chosen because neutron scattering on iodine is subdominant to that on carbon and fluorine while Coulomb interactions favor iodine. Unlike other detector technologies, a bubble chamber is nearly immune to electron signals produced by Coulomb scattering of charged pions, which makes such a measurement feasible. To mitigate multiple Coulomb scatter background with additional nuclei, a small bubble chamber 10 mm in diameter was used in this experiment, and control measurements without a target and with polytetrafluoroethylene, quartz, graphite, or crystalline iodine were carried out. With a silicon pixel telescope to track the pion trajectory, the authors successfully collected a few hundred single bubble events and measured the iodine recoil-induced nucleation threshold to be near 10 keV.

### 4.3. Neutron Sources and Capabilities

Various nuclear reactions generate neutrons. These processes can happen spontaneously, such as fission of heavy radionuclides, or be induced by impacts of energetic particles on certain target nuclei. After their production, neutrons can be further moderated or filtered to create a more desired energy profile. This section briefly summarizes the characteristics of the commonly used neutron sources as compiled in **Table 1**.

**4.3.1. Radioisotopic neutron sources.** Spontaneous fission or neutron-induced fission of heavy radionuclides produces fast neutrons in the MeV energy region. A fission reactor is capable of producing $10^{16}$ n/s for 1 MW of thermal power, of which $10^{12}$ n/s may be delivered outside the reactor. The initial neutron energy approximately follows a Maxwellian distribution with a mean energy of ~2 MeV, but it is usually moderated to thermal energies in thermal reactors. Fission neutrons may be delivered in a fast neutron beam in research reactors. The most widely accessible fission neutron source in laboratories is $^{252}$Cf, which produces ~$10^3$ n/s/μCi with a similar Maxwellian neutron energy spectrum.

Alpha bombardments on some light elements can produce MeV neutrons (α-n). Although neutrons produced by monoenergetic alphas carry only a small spread depending on the recoil angle of the daughter nuclei, α-n neutrons in reality often exhibit a broad energy distribution due to the substantial energy loss before the alphas interact with the target. As a result, the emitted neutron energy depends on both the α-decay energy and the source composition. The neutron energy spectrum will be further broadened if the α-emitter produces more than one α energy or the daughter nucleus has multiple excited energy levels. The most popular α-n neutron sources include AmBe, PuBe, and AmLi. Typical α-n neutron energies are at the MeV level and the neutron yield is usually at $10^{-6}$–$10^{-5}$ n/α in commonly used targets (99). Similarly, high-energy gammas can disintegrate light nuclei and cause neutron emission (γ-n). Common γ-n sources include Sb–Be and Y–Be. The γ-n yield is similar to that of α-n, but the produced neutrons can be nearly monoenergetic thanks to the large mean free path of high-energy gammas in the target. However, large amounts of shielding materials are often used to shield the intense gamma emission from $\gamma$-n sources, which can spoil the monoenergetic nature of the emitted neutrons.

**Table 1  Approximate properties of commonly used neutron sources**

| Source | Energy | | Yield | Timing |
|---|---|---|---|---|
| | Range (MeV) | Distribution | | |
| $^{252}$Cf | 0–10 (average of 2) | Continuous | $10^3$ n/s/μCi | $\gamma$-Tagging[a] |
| Fission reactors | 0–10 (average of 2) | Continuous | $10^{12}$–$10^{16}$ n/s/MW$_{th}$ | NA |
| AmBe | 0–10 | Continuous | $\sim 5 \times 10^{-5}$ n/α | $\gamma$-Tagging[a] |
| PuBe | 0–10 | Continuous | $\sim 5 \times 10^{-5}$ n/α | $\gamma$-Tagging[a] |
| AmLi | 0–1.5 (average of 0.45) | Continuous | $\sim 10^{-6}$ n/α | ND |
| SbBe | 0.023 | Monoenergetic | $\sim 10^{-5}$ n/γ | NA |
| YBe | 0.152 | Monoenergetic | $\sim 10^{-5}$ n/γ | NA |
| D-D | 2–3 | Monoenergetic | $\lesssim 10^9$ n/s | $\lesssim 10$ μs |
| D-T | 13–15 | Monoenergetic | $\lesssim 10^{10}$ n/s | $\lesssim 10$ μs |
| p-Li | 0–2 | Monoenergetic | Varies[b] | $\gtrsim 1$ ns |
| p-V | 0–0.2 | Monoenergetic | Varies[b] | $\gtrsim 1$ ns |

[a]Some sources emit neutrons and gammas simultaneously so the neutron timing can be inferred by tagging the gamma rays.
[b]The neutron rates of accelerator-based p-Li/V interactions strongly depend on the beam current and the target property.
Abbreviations: D, deuterium; NA, not available; ND, no data; p-Li, proton-lithium; T, tritium.

$^{252}$Cf, α-n, and γ-n sources can be made compact and are convenient to use for in situ calibrations of large detectors. However, the neutron energy spreads, and relatively low intensities limit their applications in precise nuclear recoil calibrations. Simultaneous gamma emission in some radioisotopic sources enables the neutrons to be time-tagged (100), which may allow the neutron energy to be evaluated at the event level using time of flight. This technique improves the applicability of the neutron source for nuclear recoil calibrations at the cost of further reduced neutron rates.

**4.3.2. Accelerator sources.** A two-body-to-two-body nuclear reaction has simple kinematics. When the reaction is initiated by a projectile impacting a target at rest, the energies of the ejectile and the residual nucleus are fully constrained by the projectile energy, the $Q$-value of the reaction, and the directions of the final products. The ejectile energy can be nearly monoenergetic at a chosen emission angle. Monoenergetic neutrons produced in such reactions have been widely used in and play a critical role in nuclear recoil calibrations.

The most commonly used accelerator neutron sources include $^2$H(d,n)$^3$He, $^3$H(d,n)$^4$He, $^7$Li(p,n)$^7$Be, and $^{51}$V(p,n)$^{51}$Cr. Although $^2$H(d,n)$^3$He $^3$H(d,n)$^4$He reactions are exothermic, the projectile needs to be accelerated to tens of keV in an electric field for it to overcome the Coulomb barrier of the target nucleus before the nuclear reaction can occur. Due to the relatively low voltage requirement, D-D and D-T neutron generators are often made to be compact and are considered portable. Commercial D-T neutron generators can produce $10^9$–$10^{10}$ n/s with neutron energies spanning between 13 MeV and 15 MeV for typically applied high voltage values (~100 kV). At the 90° emission angle, the neutron energy is approximately 14.1 MeV with little dependence on the acceleration high voltage. For D-D neutron generators, up to $10^7$–$10^9$ n/s can be produced, with an approximate energy of 2.45 MeV at the 90° emission angle. The energy increases to ∼3 MeV at small angles and decreases to ∼2 MeV at large angles. T-T fusion produces two neutrons with a continuous energy spectrum and is not as widely used as D-D or D-T.

The $^7$Li(p,n)$^7$Be reaction is endothermic with a $Q$-value of −1.64 MeV, so protons need to be accelerated to 1.88 MeV in the laboratory frame to react with an at-rest Li nucleus. Usually a thin film of metal lithium or LiF is evaporated on a metal backing, which is mounted at the end of a proton beam, with p-Li reactions producing neutrons and the target backing absorbing surviving protons (tantalum is usually used as the backing to minimize gamma emission from proton absorption). This reaction is capable of producing quasi-monoenergetic neutrons between a few tens of keV and a few MeV; at higher energies the cross section falls off quickly, rendering it less useful. The energy spread of the produced neutrons is determined by the straggling and energy dissipation of protons within the target; for the same initial proton energy a thin Li target leads to a narrow neutron energy distribution and a thick one yields more neutrons per unit beam current. Typically $10^6$–$10^8$ n/s can be produced with a relatively thin target. The $^{51}$V(p,n)$^{51}$Cr reaction cross section is smaller than that of $^7$Li(p,n)$^7$Be, but the near-threshold resonances in the cross section can be used to produce narrowly peaked neutrons in the keV energy region (101). Otherwise the characteristics are similar.

A distinct advantage of accelerator-based neutron sources is that they may be operated in pulsed mode, which enables the neutron production time to be measured. Large sophisticated accelerators can achieve a timing resolution of a few nanoseconds, which is sufficient for precise nTOF calculations. Commercial D-D and D-T neutron generators can also be pulsed, with a typical pulse width of microseconds or longer, although faster timing down to nanoseconds has also been demonstrated in laboratory settings (102). In addition to suppressing backgrounds using nTOF, accurate timing information from a neutron source also enables the neutron energy to be evaluated in situ or even to be monitored continuously throughout calibrations, which can greatly reduce measurement uncertainties related to the neutron energy.

### 4.3.3. Moderated/filtered neutron sources.

Fast neutrons can be moderated to lower energies using materials with high concentrations of low-$Z$ elements including hydrogen and deuterium. The most commonly used thermal neutron sources include moderated fission reactor neutrons and moderated D-D/D-T neutrons. Both Jones & Kraner (93) and Collar et al. (103) used thermalized reactor neutrons for sub-keV $^{73}$Ge recoil calibrations. With well-controlled single scatter moderation, D/H-reflected neutrons from a monoenergetic source can be shifted to a lower energy of a selected value while largely preserving the narrow energy distribution (89).

Neutrons from broadband sources may be filtered to become quasi-monoenergetic. This idea takes advantage of deep minima in the total neutron interaction cross sections of certain isotopes of iron, nickel, vanadium, manganese, and so on (104). Neutrons with energies within these deep narrow cross-section notches have high transmission in these filter materials while neutrons of other energies are strongly attenuated (filtered out). Filtered neutrons at 24 keV and 70 keV have been experimentally demonstrated using iron filters (105, 106). Multiple filter materials can be combined to improve the beam purity.

## 4.4. Survey of Published Measurement Results

In this section we present a survey of nuclear recoil measurement results over a representative set of detector technologies (scintillation, ionization, scintillation and ionization, scintillation or ionization with heat) and calibration techniques. We note that this is only a partial list of existing measurements, and more efforts exist in the literature.

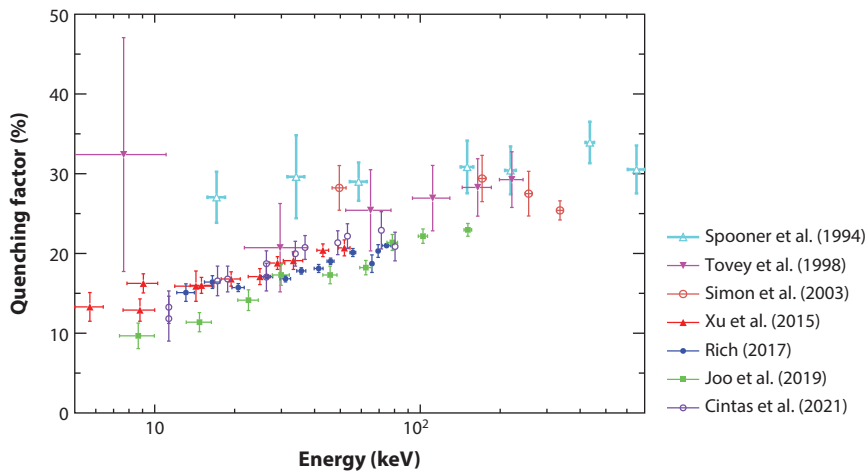

The NaI(Tl) scintillation efficiency for sodium recoils relative to that for gammas reported by different
authors (107–113). All these results were obtained with angle-tagged elastic scatters of quasi-monoenergetic
neutrons, while the DAMA/LIBRA calibration used continuous $^{252}$Cf neutron irradiation and assumed an
energy-independent quenching effect (117). Some discrepancies between measurements may be attributed
to the nonlinearity of NaI(Tl) scintillation efficiency for gamma/betas and the different reference gamma
energies used in these efforts: 100 keV in Spooner et al. (107), 60–511 keV in Simon et al. (109), 60 keV in
Tovey et al. (108), Rich (111), and Joo et al. (112), and 57 keV in Xu et al. (110) and Cintas et al. (113). The
results in References 114–116 cannot be compared directly to results in this figure due to the use of multiple
gamma/beta reference energies to account for the NaI(Tl) scintillation nonlinearity.

Due to the high light yield and the good match of emission wavelengths to photomultiplier
tube sensitivities, NaI(Tl) crystals have long been used in radiation detection and are an important target for dark matter and CEνNS detection experiments. Most notably, the DAMA/LIBRA
experiment, using an array of low-background NaI(Tl) scintillation detectors, maintains the sole
outstanding dark matter detection claim, which has been contested by several other dark matter
experiments using various target materials. Over the decades, several groups have experimentally
studied the scintillation efficiency of sodium and iodine recoils in NaI(Tl) (107–116) that can be
used to evaluate the compatibility of the DAMA/LIBRA signal with a dark matter interaction interpretation. **Figure 2** compiles the sodium recoil quenching factors (scintillation efficiency of
nuclear recoils relative to that of gammas) reported by some recent calibration efforts.

While the DAMA/LIBRA experiment used an energy-independent quenching factor of 0.3
for sodium recoils based on a continuous recoil spectrum measured with a $^{252}$Cf source (117),
all measurements included in **Figure 2** studied the energy dependence with angle-tagged elastic
neutron scatters. Generally speaking, new calibrations using lower-energy neutrons report smaller
uncertainties in the low-recoil-energy region, and the use of fast signal readouts further improves
the nTOF-based background suppression. Xu et al.'s (110) and Rich's (111) independent works
yield consistent results and comparable uncertainty levels as a result of the similar measurement
strategies employed. It is worth noting that Reference 108 reports an increased quenching factor
for the lowest energy points in the measurements, which is absent in other data. This feature may
be explained as a loss of signal efficiency at the lowest measured energies that caused the average observed energy of the recoil spectra to shift up; since signal efficiency is not discussed in
Reference 108, we cannot test this hypothesis. As demonstrated by Cintas et al. (113), due to the
nonlinearity of the NaI(Tl) scintillation response, the choice of reference gamma energy makes a significant difference in the reported quenching factor values, which correspond to different scaling factors for results shown in **Figure 2**. Measurements reported in References 114–116 compare sodium recoil-induced scintillation to that of gammas/betas at the same energy to account for this nonlinearity, and thus the results cannot be directly compared with those shown in **Figure 2**.

Semiconductor materials including silicon and germanium have low-band-gap energies and are the most widely used targets in ionization detectors. These detectors may be instrumented to also perform heat measurements or to use the heat channel to amplify low-energy ionization signals with the Neganov–Trofimov–Luke effect (31). Substantial effort has been spent in measuring the nuclear recoil ionization yields in silicon and germanium, a subset of which is summarized in **Figure 3** in comparison with the Lindhard model (52) and the model from Sarkis et al. (118).

A variety of calibration methods were used in producing these results. Gerbier et al. (121), Baudis et al. (126), Barbeau et al. (14), and IMPACT (12) used elastic scatters of quasi-monoenergetic neutrons, and Tiffenberg and colleagues (122) used a broad spectrum neutron source with event-by-event neutron energy measurement. Both techniques provide multiple

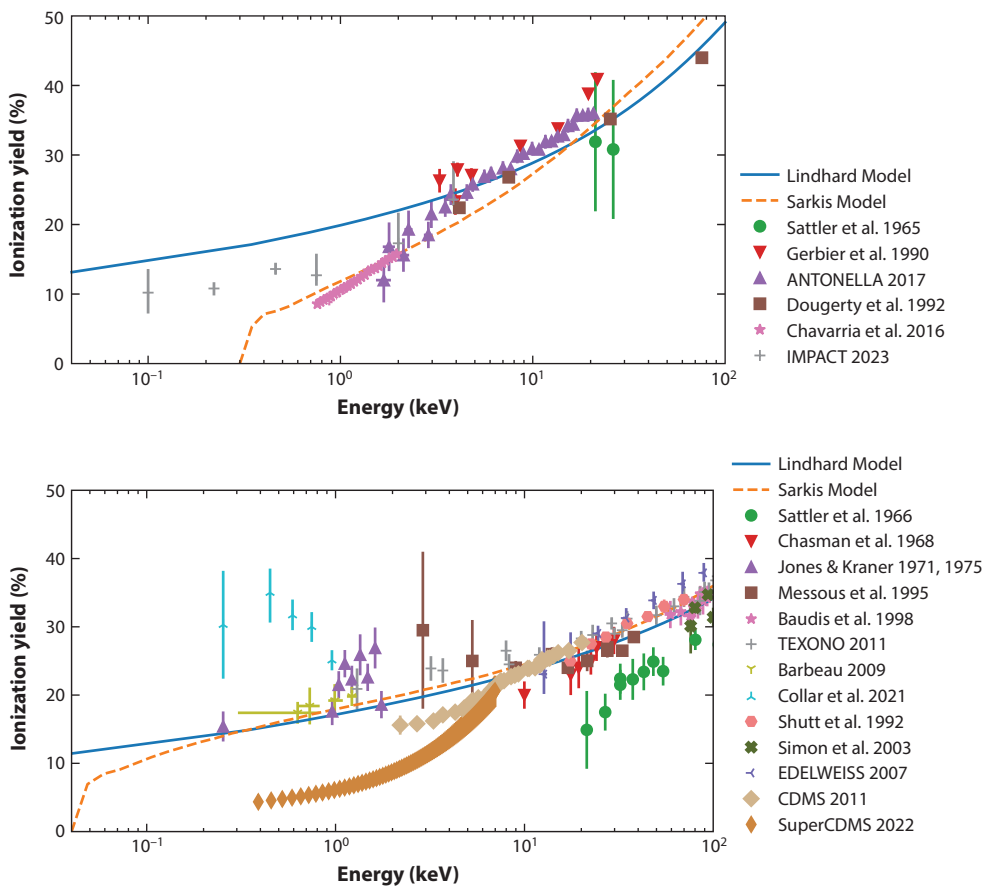

**Figure 3**

Ionization yields of Si (*top*) and Ge (*bottom*) measured by different experiments (11–16, 72, 92, 93, 119–129, 141). Predictions by Lindhard et al. (52) and Sarkis et al. (118) are also included for comparison. ANTONELLA data from Reference 122. IMPACT data from Reference 12. SuperCDMS data from Reference 15. TEXONO data from Reference 127.

independent energy points on the basis of tagged neutron angles, while the results with filtered neutrons may be subject to additional uncertainty from the neutron source purity (106). Chavarria et al. (11) and the SuperCDMS collaboration (15) used γ-n sources to probe low-energy recoils, but the lack of scatter angle information caused the results to be correlated between different recoil energies. The end-point measurements demonstrated by Sattler and colleagues (119, 124) have reduced dependence on modeling at the cost of extracting only one energy point per neutron source configuration. Neutron activation measurements by Chasman et al. (16) and Jones & Kraner (92, 93) provide sensitivity to extremely low-energy recoils, though this method is limited to the few energy points available to probe. Notably, Shutt et al. (128) and the CDMS experiment (72) carried out simultaneous measurement of charge and phonon in the high-energy region with cryogenic bolometers, which allow the ionization yield to be derived on a per-event basis.

Results above ~10 keV generally agree and are consistent with the models, while in low-energy regions, significant discrepancies between experiments emerge, and most data also deviate from the Lindhard predictions. The Sarkis model describes the data better by revisiting some of the Lindhard assumptions that may not be suitable for low-energy recoils. Recent measurements have started probing extremely low-energy regions, but the significant discrepancies between different results below 1 keV in silicon and below 5 keV in germanium call for further theoretical and experimental investigations.

Noble liquid detectors can collect both scintillation and ionization signals. When combined, these two signal channels provide accurate energy estimation (54, 130) and three-dimensional position sensitivities (131) for each recorded event. In particular, the scintillation-to-ionization signal ratio measured in dual-phase xenon TPCs demonstrates useful electron recoil–nuclear recoil discrimination power over a large energy window (down to ~1 keV). Combining the position sensitivity, background rejection capability, and low intrinsic radioactive background levels in liquid xenon, these detectors have been leading the searches for direct dark matter interactions (30, 75–77, 132, 133).

**Figure 4** shows a selection of scintillation and ionization yields of xenon recoils measured at discrete energy values between 0.3 keV and 425 keV using elastic neutron scatters. Measurements that fit continuous nuclear recoil spectra to Monte Carlo simulations, such as the ones reported in References 134 and 135, carry method-dependent systematic uncertainties and cannot be conveniently compared with energy-specific results. Because xenon has four stable isotopes with >10% natural abundance, the measurement results are the average responses of all isotopes. Among these experiments, xenon recoils between 1 and 100 keV were mostly produced by MeV-energy neutrons, and those above 100 keV were generated by 14.1 MeV D-T fusion neutrons due to the high atomic mass of xenon (136). Production of xenon recoils in the sub-keV energy region requires the use of sub-MeV neutrons or the tagging of small angle scatters. Lenardo et al. (18, 137) used a 579 keV pulsed neutron beam and tagged the scattered neutrons with separate neutron detectors to obtain recoils down to 0.3 keV; the LUX measurement used double scatters of 2.5 MeV D-D neutrons in liquid xenon and took advantage of the fine position resolution of the LUX detector for energy calculation (90). Results reported in References 18 and 90 are consistent with each other within the quoted uncertainties, while the low-energy neutron experiment reports smaller uncertainties and lower-energy measurements. LUX has recently reported a lower-energy xenon recoil measurement using single scatter events produced by a pulsed D-D neutron source (19), but due to the continuous nature of the recoil spectrum the results are not included in **Figure 4**.

Neither the scintillation nor the ionization yield of xenon recoils can be described with simple models due to the complex energy division between scintillation, ionization, and heat.

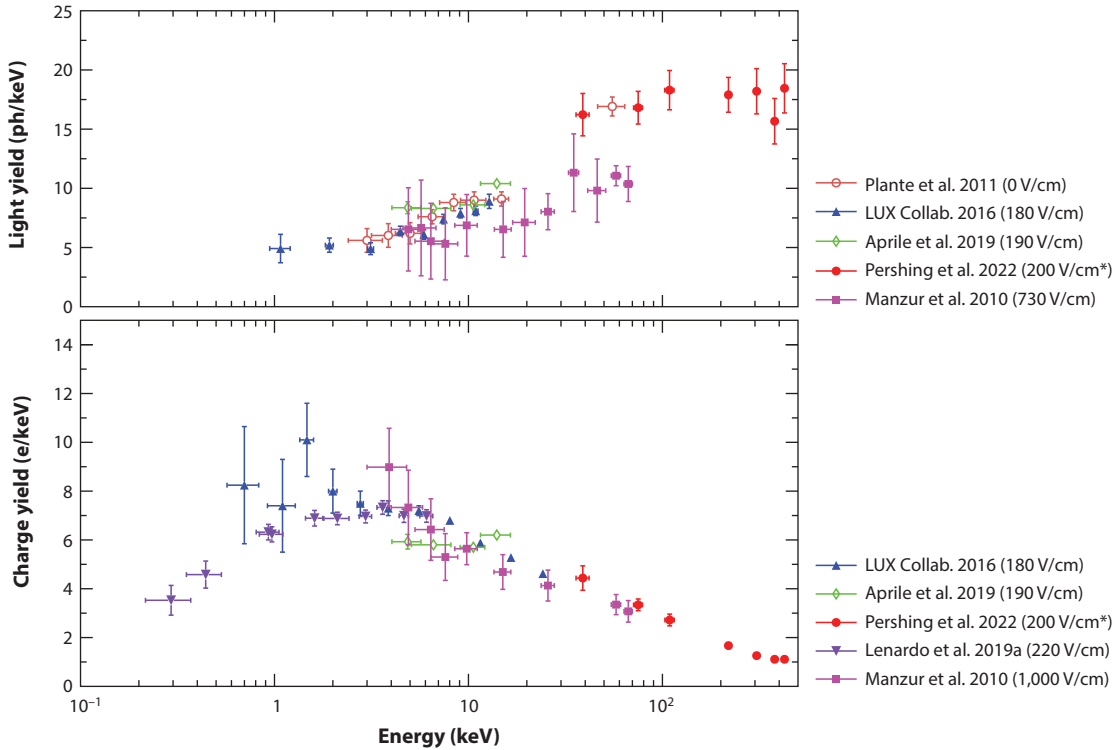

**Figure 4**

The scintillation (*top*) and ionization (*bottom*) yield values of nuclear recoils in liquid xenon measured at various recoil energy and electric field values (18, 90, 136–140). Only discrete energy calibration results are included in this plot, and for experiments measuring yields at multiple electric fields, only the lowest field results are shown for clarity. The measurement by Pershing et al. (136) uses an electric field of 200 V/cm for xenon recoil energies up to 308 keV and the results at higher energies are averaged over 3 measurements (200 V/cm, 500 V/cm, and 2,000 V/cm).

As demonstrated by Sorensen et al. (54), the summation of electronic signal channels above a few keV approximately follows the Lindhard prediction. The split of yield values between scintillation and ionization depends on both the energy of the recoils and the electric fields applied to the interaction site. Early xenon recoil measurements adopted a quenching factor–like parameter called $L_{eff}$ that compares measured scintillation outputs of nuclear recoils to those of reference gammas and varies empirically with energy and field (142, 143). The Noble Element Simulation Technique model has since been developed to predict both the light and charge yields of xenon and their spreads on the basis of fits to existing data (130, 144, 145).

Cryogenic bolometers measure the heat produced by particle interactions; electronic signals including scintillation and ionization, when not separately detected, often contribute to the total heat observed. As a result, conventional bolometers may not have a strong need to separately calibrate the energy scale of the detectors for nuclear recoils. However, the fraction of total energy dissipated in the electronic channels offers a means for nuclear recoil signals to be distinguished from electron recoil backgrounds that produce little heat and from excessive low-energy noise events that appear to be heat only (85). Some bolometer experiments have developed dual readouts with phonon and charge or phonon and light, and their ionization or scintillation yields have to be separately calibrated. The in situ germanium ionization yield measurement by SuperCDMS (72) is included in **Figure 3**, and the CRESST experiment has calibrated the scintillation yield of $CaWO_4$ crystals operated in the bolometer mode (79). Recently the nuclear recoil response of superfluid helium was also studied (146).

## 4.5. Mitigation of Calibration Biases

Biases in a calibration experiment can cause the evaluated nuclear recoil signal strength to deviate from true values. As discussed in Section 4.4, a possible source of bias is inadequately evaluated signal efficiency in the target detector. A finite trigger threshold in an experiment, by definition, causes events below the threshold to not get recorded or recorded with reduced efficiencies. This loss of low-energy signals, when uncorrected for, biases the energy distribution of recorded events toward higher values, which can be incorrectly interpreted as a higher-than-real signal yield near the detector threshold. We recommend that all nuclear recoil calibration experiments rigorously evaluate the detector's signal acceptance as a function of observable energy and include a dedicated discussion on signal efficiency in the publications. When interaction cross-section data exist, the measured event rate can also be compared to that expected as an additional check of signal acceptance. One method to mitigate this trigger bias is to adopt a trigger scheme that is independent of the energy deposition in the target detector. In coincidence calibrations that tag neutrons, the neutron timing obtained from the neutron source or from the tagging detectors can enable a triggerless data acquisition in which the data recording is informed by the neutron timing rather than by the detector response. As a result, no additional signal loss would occur for low-energy signals. This is especially important for the detection of extremely low-energy signals that can produce zero observables with a nonnegligible probability. This detector-agnostic trigger scheme will increase the recorded data volume and requires the experiment to properly handle accidental coincidence backgrounds, so its implementation can be a challenge and may not be suitable for all experiments.

Similar efficiency-related biases can originate from analyses. For example, analysis cuts that target saturated signals or distorted pulses can disproportionately reject high-energy signals and cause the remaining event distribution to shift toward lower energies. Well-intended cuts in an analysis can also bias signal acceptance in more subtle ways than the example above, especially if the signal statistics are low. Therefore, signal efficiencies should be evaluated in all energy regions of interest. The idea of a blind analysis has been proposed and tested to mitigate both explicit and implicit biases. In such an analysis, all data selection cuts are thoroughly vetted using simulation or a small portion of data, and once the cuts are finalized and applied to the analysis data set, no more changes to the cuts are allowed. Analyses conducted in this manner are less likely to be biased, but when this method is not implemented correctly it can suffer from unexpected background contamination and introduce a different form of inaccuracy.

Another source of calibration bias can rise from the measurement of neutron scatters at angles where the interaction cross section undergoes substantial changes. For example, elastic neutron scatters often have a forward-peaked cross section, and as a result the average neutron scatter angle for events tagged by a neutron coincidence detector will be lower than that calculated from the centers of the detector positions. This can cause the calculated nuclear recoil energy to be overestimated and the derived signal yield to be underestimated. This issue may be mitigated by comparing data with simulation that incorporates the angular cross-section dependence, and uncertainties in the cross-section data should be propagated into the final results. This is also true for neutron scatters near resonances where the interaction cross sections change drastically with scatter angle or energy. Such scenarios should be avoided if possible.

Lastly, for the calibration of detectors with slow responses or recoil events producing small signals, the choice of event window in the target detector can lead to inconsistencies in obtained results. An integration time window that is too short captures only part of the signal while one that is too long may get contaminated by noise pulses. In addition, for signals consisting of only a small number of pulses, the arrival time of the first pulse with respect to the interaction time carries significant uncertainty, and as a result event windows initiated by the first detected pulse vary from event to event. In these cases, the event window should be defined by the actual interaction time informed by the neutron production time or detection time in a tagging detector to ensure consistency. The choice of event integration window shall also be reported.

## 4.6. Presentation of Experimental Results

Nuclear recoil calibration results can be presented in one of the three following ways:

- number of quanta, the measured number of quanta such as photons or electrons at a given energy;
- yield value, the measured number of quanta normalized to unit energy; and
- quenching factor, the measured yield value relative to that of reference gammas or electron recoils of a specific energy.

The choice of an experiment can be driven by the direct application of the calibration, the availability of detector information, or the choices of past efforts. Surveyed results in Section 4.4 mostly report quenching factors or yield values. The lack of a consistent convention has prevented some calibration results from being meaningfully incorporated in comparisons to similar efforts or applications in dark matter or neutrino searches.

We recommend publishing the measured number of quanta with signal efficiencies evaluated and corrected for, when available, in tabulated forms to facilitate the comparison and modeling of results from different calibration efforts. The number of quanta is usually a direct product of a measurement and carries uncertainties (statistics, single quanta response, signal collection efficiency, etc.) that are nearly orthogonal to those of the estimated recoil energy (neutron energy, scatter angle, etc.). On the other hand, the derived yield values could be affected by all the aforementioned uncertainties, and the energy-associated uncertainties may be overrepresented in a yield-energy presentation, which weakens the power of comparison between experiments.

The energy-normalized yield values and their dependence on energy contain important information on the microphysics governing the particle's interactions with the medium. Therefore, it is appropriate for the yield values to be presented in graphical forms, especially when the yields are compared to model predictions. A graph of measured quanta, in contrast, is usually dominated by its dependence on the recoil energy and may not well illustrate the property of the target material. Generally we recommend the graphical presentation of yields and numerical report of quanta numbers. An exception is for experiments that derive yield values using nuclear recoils with a substantial energy spread where the quanta number is difficult to define and evaluate and can carry large uncertainties, but the yield can be insensitive to energy and have reduced uncertainties.

Certain measurements cannot obtain the absolute numbers of quanta or calculate the absolute yields. This can happen when the detector gain is too low for the single quanta response to be resolved with high accuracy or the signal detection efficiency cannot be reliably evaluated. In these cases, it is useful to present the experimental results in the form of quenching factors by normalizing the detector response to nuclear recoils to that of reference gammas (or betas) that share the same single quanta responses and detector efficiencies. In this process, the choice of reference radiation type and energy becomes crucial. First, the reference interactions should resemble the energy and position distributions of the nuclear recoils for the unmeasured detector factors to correctly cancel out. Second, the exact type of interaction should be convenient to produce in other experiments, including those of large scales that may be well shielded and not easily accessible, for the quenching factor to be practically used. In the case of NaI(Tl) measurements, Gerbier et al. (114), Chagani et al. (115), and Collar (116) reported the quenching factors normalized to gamma/beta at the same energy as the nuclear recoils, which is well motivated from the perspective of physics understanding, but if the assumed gamma/beta yield dependence on energy is not reported, these results can be inconvenient to use in another experiment for application or comparison. If necessary, a quenching factor measurement can report the results using multiple reference sources, as demonstrated by Cintas et al. (113).

The presentation of experimental uncertainties will also benefit from a convention. Statistical uncertainties impact a measurement in a relatively well-defined manner, but systematic uncertainties are unique to each experiment and should be reported separately. For example, the uncertainty in an elastic neutron scatter calibration can get complex when multiple neutron tagging detectors are used. In this case, the uncertainties from the neutron source, such as energy and timing, affect all coincidence angles similarly. However, uncertainties from the neutron beam alignment and the target detector position can produce correlated or anticorrelated uncertainties in events tagged with different neutron detectors. The separation of different sources of uncertainties and a dedicated discussion should always be included in a publication.

## 4.7. Future Measurements and Challenges

Searches for low-mass dark matter interactions and reactor CE$\nu$NS have driven nuclear recoil calibrations to the sub-keV regime. To produce elastic neutron interactions at these energies, neutrons with tens of keV of energy or lower are needed, but fast tagging of these neutrons has been a challenge. Low-energy neutrons may not deposit enough prompt energy in a tagging detector to be detected or to be differentiated from gamma interactions. This problem may be mitigated with neutron-capture detectors, but the long capture time can cause accidental coincidence backgrounds to increase. Alternative calibration methods using gamma interactions or inelastic neutron scatters could avoid these difficulties but have not demonstrated the required accuracy. In addition, as the signal energy approaches the quantum limit, nuclear recoil events can produce zero observables with a significant probability (such as the zero component of a Poisson distribution), so the missing signals and their implications for the evaluated detector response need to be carefully studied. Further, the spread of low-energy quantized signals does not necessarily follow simple Poisson or Gaussian distributions, which makes modeling of the signals and data-model comparisons extremely difficult. Therefore, new methods to improve low-energy calibration techniques need to be developed.

The need to detect extremely low-energy nuclear recoils has led to renewed interest in the Migdal effect (60). As proposed by Migdal and reformulated by Ibe et al. (61), a nuclear recoil interaction displaces the nucleus relative to shell electrons, which results in a finite probability for the atom to be ionized. This process causes a small fraction of nuclear recoil events to contain an electron recoil component that can be at the keV level or above; both the enhanced signal energy and the better detectability of electron recoils enable a large class of particle detectors to substantially lower their effective energy thresholds for nuclear recoils. Although it has not been experimentally observed, several experiments have studied the potential sensitivity gains with this effect (29, 63–65). Verification of the Migdal effect may be accomplished with techniques similar to those used in nuclear recoil calibrations, except that the expected signal in the target detector becomes an electron recoil (for subthreshold nuclear recoils) or a combination of electron recoil and nuclear recoil (for higher-energy nuclear recoils). A definitive measurement of the Migdal effect requires the target detector to be capable of resolving the nuclear recoil and electron recoil tracks (147) or to have sufficient nuclear recoil–electron recoil discrimination capabilities. Several experimental efforts to study the Migdal effects are underway.

Besides the low-energy thrust, WIMP dark matter experiments are also expanding their searches to include nuclear recoil signals in high-energy regions up to hundreds of keVs. This is partially motivated by the absence of positive signals in the low-energy region despite fast improvements in experimental sensitivity; at the same time, well-motivated theories, including EFT (24) and inelastic dark matter interaction models (148), predict that signals could primarily be present in high-energy regions. For heavy elements such as xenon and germanium, neutrons of tens of MeV are needed to produce recoils above 100 keV. Compared with lower-energy neutrons, high-energy neutron sources suitable for calibrations are less available, and they also require substantially more shielding materials, which can produce significant secondary neutrons and gammas and contaminate the experiment. In addition, these neutrons are capable of exciting more nuclear energy levels or even causing spallation in heavy targets, so their elastic scatter cross sections may be relatively suppressed. Continued research and development is needed for the calibration of dark matter detectors to exploit new sensitivities that can be enabled by high-energy recoil searches.

New advancements in particle physics can be enabled by improvements in calibration precision. For example, if a CEνNS experiment measures the total interaction rate at the percent level, it can constrain the weak mixing angle at low $Q$-values more accurately than the current best results, and a deviation of the measured CEνNS energy spectrum from that predicted by the SM may indicate a nonzero neutrino magnetic moment or other nonstandard interactions (41). However, such measurements require the nuclear recoil signal yields to be calibrated at the subpercent level. Calibrations of such high precision at the keV level have not been achieved and will likely require the development of new detector technologies and new calibration techniques.

As dark matter and CEνNS detectors get bigger and more complex, the collected detector signals could be strongly shaped by detector-specific conditions, such as the detector geometry, purity of the target, signal collection efficiency, and so on, in addition to target material properties. As a result, the calibration results from one setup may not be easily translated into responses in a similar detector. For this reason, there is a strong need for in situ calibrations where nuclear recoil calibrations are conducted with the very detector in the same environment where searches for dark matter or CEνNS are carried out. However, these experiments are often operated in underground laboratories or other locations with limited space and access, so the types of neutron sources suitable for in situ calibrations are limited. Portable neutron sources such as D-D/D-T neutron generators, photoneutrons, or α-n emitters are often used in these efforts. Although these calibrations may not produce target responses as accurate as small-scale efforts at sophisticated facilities, they can provide overall detector responses for signal evaluations.

Finally, due to the similar nature of WIMP dark matter signals and CEνNS events, CEνNS signals from solar and atmospheric neutrinos can show up in a WIMP search experiment. These irreducible CEνNS signals will obscure WIMP signals if the WIMP interaction rate is below that of CEνNS, and as a result large detectors' expected sensitivities to WIMPs could diminish when they approach this neutrino floor. One possible strategy to mitigate this background is to leverage the different directions of incoming neutrinos and dark matter particles, which produce nuclear recoils of different angular distributions. Low-density detectors such as those using gas targets can have sensitivity to tracks (149), and nuclear recoil calibrations with gas targets have been demonstrated (150). It has also been suggested that recoils in a TPC may have different yield values for different track orientations with respect to the applied electric field, but this effect remains to be verified. Since the neutron kinematics (Equation 5) constrain the nuclear recoil direction, the calibration techniques discussed in this section are also applicable to directionality studies of these nuclear recoil detector technologies.

## 5. CLOSING REMARKS

New particle physics discoveries may be enabled by experimental searches for low-energy nuclear recoil signals produced by WIMP dark matter or neutrino scattering interactions. Of all challenges facing these experiments, the detection and calibration of nuclear recoils in the relevant energy region are among the most important. A wealth of experimental techniques have been explored in past efforts. By reviewing what has been achieved experimentally and by highlighting the underlying considerations shaping these experiments, we aim to provide useful guidance to future endeavors in this direction.

For the main nuclear recoil detection techniques that are currently being investigated, we have observed a fast increase in calibration accuracy and a substantial reduction of energy thresholds in recent years. In particular, significant progress has been made with the elastic neutron scatter method that uses (quasi-)monoenergetic neutrons and tags their scatter angles off target nuclei. Combining the nTOF technique and PID methods has enabled nearly background-free calibrations down to energy regions where only a handful of quanta are produced. Despite its demonstrated success, improvements to this method and the development of alternative methods are needed to meet the increasing calibration demand of more sensitive dark matter and neutrino experiments, especially for efforts focusing on low-energy signals or those requiring high precision.

In situ calibrations that produce broad spectra of nuclear recoils in rare event search experiments still play an irreplaceable role in validating the detectors' signal acceptance. Such data can be used to guard against possible biases in both data collection and analysis. A detector's overall response to nuclear recoil signals is a convolution of the properties of the target material and the specific detector configuration that controls the signal production and collection processes, so in situ calibrations are naturally complementary to ex situ calibrations that focus on material properties.

Lastly, all experimental efforts are susceptible to uncertainties and biases. We summarize the common sources of biases in calibration efforts and propose corresponding mitigation strategies. For different experimental efforts to contribute to a coherent understanding of how a specific detector technology responds to signals, it is critical for a convention on data presentation to be developed and followed so ambiguity in interexperiment data comparisons can be minimized. We propose such a convention in Section 4.6. After all, the possibility of the independent examination and validation of experimental results is a fundamental principle of scientific research.

## DISCLOSURE STATEMENT

The authors are not aware of any affiliations, memberships, funding, or financial holdings that might be perceived as affecting the objectivity of this review.

## ACKNOWLEDGMENTS

J.X. is supported by the US Department of Energy (DOE) Office of Science, Office of High Energy Physics under work proposal SCW1676 awarded to the Lawrence Livermore National Laboratory (LLNL). LLNL is operated by Lawrence Livermore National Security, LLC, for the DOE, National Nuclear Security Administration under contract DE-AC52-07NA27344. This manuscript has been approved by LLNL for public release with an identification number of LLNL-JRNL-842240. Z.H. is supported by the Natural Sciences and Engineering Research Council of Canada (SAPPJ-2022-00034.

Annu. Rev. Nucl. Part. Sci. 2023.73:95-121. Downloaded from www.annualreviews.org
Access provided by 98.51.119.174 on 10/28/23. See copyright for approved use.

We thank Adam Bernstein, Tenzing Joshi, Brian Lenardo, Rachel Mannino, Ethan Bernard, Teal Pershing, and Charles Prior for their helpful feedback on early versions of this article. We thank Vetri Velan, Matthew Szydagis, Hugh Lippincott, Shawn Westerdale, Duncan Adams, Nathaniel Bowden, Viacheslav Aleksandrovich Li, and Tarek Saab for their valuable technical input on some topics covered in this review.

## LITERATURE CITED

1. Gaitskell RJ. *Annu. Rev. Nucl. Part. Sci.* 54:315 (2004)
2. Buchmüller W, Peccei R, Yanagida T. *Annu. Rev. Nucl. Part. Sci.* 55:311 (2005)
3. Marciano WJ, Parsa Z. *Annu. Rev. Nucl. Part. Sci.* 36:171 (1986)
4. Freedman DZ. *Phys. Rev. D* 9:1389 (1974)
5. Akimov D, et al. *Science* 357(6356):1123 (2017)
6. Lewin JD, Smith PF. *Astropart. Phys.* 6:87 (1996)
7. Lindhard J, Scharff M. *Phys. Rev.* 124(1):128 (1961)
8. Essig R, et al. arXiv:2203.08297 [hep-ph] (2022)
9. Akindele OA, et al. arXiv:2203.07214 [hep-ex] (2022)
10. Bowen M, Huber P. *Phys. Rev. D* 102(5):053008 (2020)
11. Chavarria AE, et al. *Phys. Rev. D* 94(8):082007 (2016)
12. Albakry MR, et al. arXiv:2303.02196 [physics.ins-det] (2023)
13. Collar JI, Kavner ARL, Lewis CM. *Phys. Rev. D* 103(12):122003 (2021)
14. Barbeau PS, Collar JI, Tench O. *J. Cosmol. Astropart. Phys.* 0709:009 (2007)
15. Albakry MF, et al. *Phys. Rev. D* 105(12):122002 (2022)
16. Chasman C, Jones KW, Kraner HW, Brandt W. *Phys. Rev. Lett.* 21(20):1430 (1968)
17. Akerib DS, et al. *Phys. Rev. Lett.* 116(16):161301 (2016)
18. Lenardo BG, et al. *Phys. Rev. Lett.* 123(23):231106 (2019a)
19. Akerib DS, et al. (LUX Collab.) arXiv:2210.05859 [physics.ins-det] (2022)
20. Aghanim N, et al. (Planck Collab.) *Astron. Astrophys.* 641:A6 (2020). Erratum. *Astron. Astrophys.* 652:C4
21. Feng JL. *Annu. Rev. Astron. Astrophys.* 48:495 (2010)
22. Jungman G, Kamionkowski M, Griest K. *Phys. Rep.* 267(5):195 (1996)
23. Steigman G, Dasgupta B, Beacom JF. *Phys. Rev. D* 86(2):023506 (2012)
24. Fitzpatrick AL, et al. *J. Cosmol. Astropart. Phys.* 1302:004 (2013)
25. Essig R, et al. arXiv:1311.0029 [hep-ph] (2013)
26. Essig R, Pradler J, Sholapurkar M, Yu TT. *Phys. Rev. Lett.* 124(2):021801 (2020)
27. Gu PH, He XG. *Phys. Lett. B* 778:292 (2018)
28. Escudero M, Berlin A, Hooper D, Lin MX. *J. Cosmol. Astropart. Phys.* 1612:029 (2016)
29. Akerib DS, et al. *Phys. Rev. Lett.* 122(13):131301 (2019)
30. Aprile E, et al. *Phys. Rev. Lett.* 123(25):251801 (2019)
31. Agnese R, et al. *Phys. Rev. Lett.* 116(7):071301 (2016)
32. Akerib DS, et al. *Phys. Rev. D* 104(6):062005 (2021)
33. Aprile E, et al. arXiv:2210.07591 [hep-ex] (2022)
34. Adhikari P, et al. *Phys. Rev. D* 102(8):082001 (2020)
35. Agnes P, et al. *Phys. Rev. D* 101(6):062002 (2020)
36. Balantekin AB, Kayser B. *Annu. Rev. Nucl. Part. Sci.* 68:313 (2018)
37. Fukuda Y, et al. *Phys. Rev. Lett.* 81(8):1562 (1998)
38. Ahmad QR, et al. *Phys. Rev. Lett.* 87(7):071301 (2001)
39. de Gouvêa A. *Annu. Rev. Nucl. Part. Sci.* 66:197 (2016)
40. KATRIN Collab. *Nat. Phys.* 18(2):160 (2022)
41. Scholberg K. *Phys. Rev. D* 73:033005 (2006)
42. Akimov D, et al. *Phys. Rev. Lett.* 126(1):012002 (2021)
43. Anderson AJ, et al. *Phys. Rev. D* 86(1):013004 (2012)
44. Janka HT, Melson T, Summa A. *Annu. Rev. Nucl. Part. Sci.* 66:341 (2016)

*118    Xu et al.*

# Contents

**Errata**

An online log of corrections to *Annual Review of Nuclear and Particle Science* articles may
be found at http://www.annualreviews.org/errata/nucl